\documentclass[12pt]{article}
 \usepackage{putex, epsfig}
 \usepackage{graphicx}
 \usepackage{placeins,verbatim}

 \newcommand{\beq}[1]{\begin{eqnarray}\label{#1}}
 \newcommand{\eeq}{\end{eqnarray}}
 \newcommand{\nwl}{\nonumber}
 \newcommand{\nwsgn}[1]{\mathrm{sgn}{#1}}

\begin{document}

 \preprint{hep-th/0708.xxxx \\ ITP-BJUT-000A}

 \institution{ITP-CAS}{Institute of Theoretical Physics, Beijing University of Technology}

 \title{Extending BCCNR flavored geometry to the negative coupling constant region}

 \authors{Ding-fang Zeng}
 \footnote{dfzeng@bjut.edu.cn}

 \abstract{
 We extend BCCNR's flavored KW
 and KS geometries into the negative coupling constant
 region and try giving physical explanations from the
 dual gauge theory side. We show that, in the extended
 region, except the coupling constant being negative,
 all other fields of the super-gravity theory having
 good properties which can be described by the dual gauge theory.
 }


 \date{\today}

 \maketitle

 \tableofcontents

 \section{Introduction}

 AdS/CFT correspondence \cite{Maldacena:1997re}\cite{GKPWitten:1998}
 is one of the most powerful
 analytical tools in the studying of strongly coupled gauge theories.
 In Maldacena's original conjecture, the CFT is typically denoted by the $\mathcal{N}=4$
 SYM, which is an exact conformal formal field theory. Since
 Maldacena's work, many other works appear and the AdS/CFT is
 generalized to more interesting non-conformal models\cite{nonCon-AdSCFT-model}, minimally or
 non-supersymmetric gauge theories \cite{{KW9807},{KN9911},{KT0002},{KS0007},{GHK0405}}. To use
 AdS/CFT studying gauge theories containing flavor quarks, reference
 \cite{KarchKatz0206} proposed a probe approximation method. This
 method introduces finite number of space-time filling flavor
 D7-branes into the background formed by the $N_c\rightarrow\infty$ color
 D3-branes and studies the flavor branes' configuration with minimal DBI action but
 neglects their back-reactions on the color background,
 for concrete examples please see references \cite{ACR0408, Ouyang0311P0506, SSK0309P0411}. In the dual field theory,
 this corresponds to considering effects of the color dynamics on
 the flavors completely but neglect the back-reaction of flavors on the
 color dynamics totally. In terminology of the lattice gauge theory, this is
 the ``quenched'' approximation.

 To go beyond this ``non-backreacting'' approximation, reference
 \cite{BLZV0406} and \cite{BKS0708} provide an exact method.
 The regret is, this method cannot give exactly
 analytical expressions for the supergravity solution. It
 depends on perturbations or numerics heavily. While on
 the basis of a serial of works, \cite{BCCKP0505,Casero:2006pt,Paredes:2006wb,MT0606,Casero:2007pz}, BCCNR
 proposed a new method of studying flavor quarks's effects on the
 color dynamics in two successive works \cite{BCCNR0612}
 \cite{BCCNR0706}. In
 reference \cite{BCCNR0612}, BCCNR considered flavoring of
 the Klebanov-Witten field theory/geometry, which is a type IIB
 solution dual to the $SU(N_c)\times SU(N_c)$ $\mathcal{N}=4$ SCFT.
 In reference \cite{BCCNR0706}, BCCNR considered flavoring of the
 Klebanov-Tseytlin/Strassler cascading gauge theories. In this two works,
 BCCNR add flavor branes of numbers comparable with that of the color D3-branes,
 and distribute them homogeneously on a two-sphere of the
 transverse space of the background space-time. By this smearing
 procedure, BCCNR obtained exact and analytical super-gravity
 description of KW, KT and KS gauge theory dynamics back-reacted
 by the flavors. To study flavor physics from the holographic perspective
 is a wide and interesting area. This includes not only the developing of
 new methods, e.g. BCCNR's smearing procedure, but also some concrete
 phenomenologies, e.g. the flavor related phase transition of
 \cite{flavorPhaseTransition}. It still includes many works which
 cannot be distinguished so clearly, e.g., the recent works of
 \cite{{AELS0610},{Sieg0704},{ARR0704}}. For complete references,
 please see the citation list of \cite{KarchKatz0206}.

 One aspect which is not fully explored by BCCNR is their flavored
 geometry's ultraviolet completion. According to their first paper
 \cite{BCCNR0612}, there is a Landau pole at some critical energy
 scale which prevent one from developing a complete UV description of
 the theory. Below this critical energy scale, the string coupling
 constant $e^{\phi}$ takes positive value. While above this critical
 value, the string coupling constant $e^\phi$ becomes negative. We
 observed that, if the string coupling constant is allowed to take
 negative values, i.e. the dilaton field contains a
 non-trivial (but constant) imaginary part, then BCCNR's flavored geometry can
 be extended to arbitrary high energy scales and a possible UV
 completion can be obtained. The next section of this paper describes
 this extension of the BCCNR flavored KW geometry and the resulting
 geometry's asymptotical behaviors. While
 the following section gives one comment about the BCCNR flavored KS
 geometry and study its extensions to the negative coupling constant
 region. The last section is our conclusions.

 \section{Extension of the BCCNR flavored KW geometry}

 KW field theory/geometry,
 is a type IIB solution dual to an $SU(N_C)$
 $\times$ $SU(N_C)$ $\mathcal{N}=1$ SCFT,
 \beq{}
 ds^2=h(r)^{-1/2}dx_{1,3}^2+h(r)^{1/2}[dr^2+r^2dT_{1,1}^2]
 \nwl\\
 h(r)=27\pi N_cg_s\alpha^{\prime2}/(4r^4)
 \nwl\\
 \ F_5=\frac{1}{g_s}(1+\star)dt\wedge dx^1\wedge dx^2\wedge dx^3\wedge
 dh(r)^{-1}
 \\
 dT_{1,1}^2=e^{\theta_12}+e^{\phi_12}+e^{\theta_22}+e^{\phi_22}+e^{\psi2}
 \nwl\\
 e^{\psi}=\frac{1}{\sqrt{9}}(d\psi+\cos\theta^1d\phi^1+\cos\theta^2d\phi^2)
 \nwl\\
 e^{\theta_{1,2}}=\frac{1}{\sqrt{6}}d\theta^{1,2},\ e^{\phi_{1,2}}=\frac{1}{\sqrt{6}}\sin\theta^{1,2}d\phi^{1,2}
 \label{KWgeometry}
 \eeq
 with constant dilaton and all the other fields in the Type IIB
 supergravity vanishing. From the string perspective, this
 background describes D3-branes placed at the tip of some conifold.
 While from the dual gauge theory perspective, this geometry
 describes a gauge field system with matter fields $A_i$, $B_i$, $i=1,2$
 transforming in the bi-fundamental representation $(N_c, \bar{N}_c)$
 and $(\bar{N}_c, N_c)$ respectively. As usual, the gauge field
 of the system transforms according to the adjoint representation.

 \subsection{Basic ingredients of the BCCNR flavored KW geometry}

 When two sets of space-time filling D7-branes ( each of number $N_f$ ) are
 added to the Klebanov-Witten geometry on hyper surfaces parameterized by
 \beq{}
 \xi_1^\alpha=\{x^0,x^1,x^2,x^3,r,\theta_2,\phi_2,\psi\},\
 \theta_1=\mathrm{const.},\ \phi_1=\mathrm{const.};
 \nwl\\
 \xi_2^\alpha=\{x^0,x^1,x^2,x^3,r,\theta_1,\phi_1,\psi\},\
 \theta_2=\mathrm{const.},\ \phi_2=\mathrm{const.};
 \label{D7embeddingWay}
 \eeq
 the resulting geometry describes a dual gauge system
 with flavor quarks \cite{ACR0408,Ouyang0311P0506}, which transforms
 in the fundamental and
 anti-fundamental representations of the gauge group.
 When $N_f$ is very large ($N_f$ $\sim$ $N_c$
 , $N_c\rightarrow\infty$), BCCNR observed that the
 two sets of D7-branes can be distributed on the
 respective transverse directions homogeneously, so that the
 resulting brane+supergravity system can be described by the
 following action
 \beq{}
 &&\hspace{-5mm}S=\frac{1}{2\kappa_{10}^2}\int d^{10}x\sqrt{-G}\left[
 R-\frac{1}{2}\partial_M\phi\partial^M\phi-\frac{1}{2}e^{2\phi}|F_1|^2
 -\frac{1}{4}|F_5|^2\right]
 \nwl\\
 &&\hspace{2mm}
 -\frac{N_fT_7}{4\pi}\left[
 \int d^{10}xe^\phi\sum_{i=1}^2\sin\theta_i\sqrt{-\hat{G}^{(i)}_8}
 -\int[\mathrm{Vol}(Y_1)+\mathrm{Vol}(Y_2)]\wedge C_8
 \right].
 \label{smearingProcedure}
 \eeq
 While the metric ansatz of system can be written as
 \beq{}
 &&\hspace{-3mm}dr=e^fd\rho
 ,\ r_{\rho\rightarrow-\infty}\rightarrow0
 \nwl\\
 &&\hspace{-3mm}ds^2=h^{-\frac{1}{2}}dx_{1,3}^2
 +h^{\frac{1}{2}}(e^{2f}d\rho^2+e^{2f}d\hat{T}_{1,1}^2)
 \nwl\\
 &&\hspace{-3mm}d\hat{T}_{1,1}^2=e^{2g-2f}[e^{\theta_12}+e^{\phi_12}+e^{\theta_22}+e^{\phi_22}]+e^{\psi2}
 \label{BCCNRmetricAnsatz}
 \eeq
 where $r$ is the usual radial coordinate like that of
 eq\eqref{KWgeometry}, $\rho$ is a new radial coordinate which plays
 the same role as $r$. For detailed ansatz of the dilaton and other
 form fields, we refer the reader to reference \cite{BCCNR0612}.

 Both the embedding scheme \eqref{D7embeddingWay} of the flavor D7-branes and
 the smearing procedure \eqref{smearingProcedure} preserve four of the
 unflavored geometry's super charges. Using this fact BCCNR set up
 the first order BPS equations satisfied by functions in the metric
 ansatz and find the corresponding solutions as
 \beq{}
 \left\{
 \begin{array}{l}
 \dot{g}=e^{2f-2g}
 \\
 \dot{f}=3-2e^{2f-2g}-\frac{3N_f}{8\pi}e^\phi
 \\
 \dot{\phi}=\frac{3N_f}{4\pi}e^\phi
 \\
 \dot{h}=-27\pi N_ce^{-4g}
 \end{array}
 \right.,\
 \left\{
 \begin{array}{l}
 e^\phi=-\frac{4\pi}{3N_f}\frac{1}{\rho},
 \\
 e^g=\left[(1-6\rho)e^{6\rho}+c_1\right]^{\frac{1}{6}}
 \\
 e^f=e^{3\rho+\frac{1}{2}\ln(-6\rho)}\left[(1-6\rho)e^{6\rho}+c_1\right]^{-\frac{1}{3}}
 \\
 h(\rho)=-27\pi N_c\int d\rho\left[(1-6\rho)e^{6\rho}+c_1\right]^{-\frac{2}{3}}
 \end{array}
 \right.
 \label{BCCNReq+sol}
 \eeq
 BCCNR stated that to assure the positivity of the
 string coupling constant $e^\phi=-4\pi/(3N_f\rho)>0$, the
 sensible value of $\rho$ should be greater than zero. In
 holographic languages, $\rho=0$ marks a specific energy
 scale. The fact that $\rho$ cannot take values greater
 than zero means that, there exists a Landau pole in the dual
 gauge theory. This Landau pole prevents us from knowing
 things about the gauge theory above this energy scale.
 However, if we allow $e^\phi$ can be less than zero,
 then we can extend BCCNR's flavored geometry into
 the $\rho>0$ region, and obtain a possible UV completion
 of BCCNR flavored KW gauge theory.

 \subsection{Extension of the BCCNR Solution}

 So let us assume that $e^\phi$ can be less than zero,
 and re-examine the solutions to the first order equations
 in the left side of \eqref{BCCNReq+sol}. For
 the dilaton field we easily get
 \beq{}
 e^{\phi}=\left\{
 \begin{array}{l}
 -\frac{4\pi}{3N_f}\frac{1}{\rho+c},\ \rho\geq0
 \\
 -\frac{4\pi}{3N_f}\frac{1}{\rho-c},\ \rho<0
 \end{array}
 \right.\equiv-\frac{4\pi}{3N_f}\frac{1}{\rho+\nwsgn{\rho}\cdot c} \label{dilatonFlavoredKW}
 \eeq
 Where $c$ is an integration constant whose physical meaning
 is, the inverse of the string coupling(modding out a constant
 coefficient) at the $\rho=0$ point. In BCCNR's first work
 \cite{BCCNR0612}, this constant is set to $0$;
 while in their second paper\cite{BCCNR0706}, a similar constant
 is not set so. It is
 worth to note that $c$ has no meaning of critical radial
 coordinate from any sense, so it should not be removed by translating
 redefinition of the $\rho$ coordinate. To avoid singularities when
 $\rho$ takes finite values, we will assume that $c>0$.
 And to let
 the super-gravity descriptions be valid when $\rho\approx0$, we will
 assume that $c>>1$,
 and simultaneously $N_cg_s\approx N_ce^\phi|_{\rho=0}>>1$.
 Obviously, according to the solution \eqref{dilatonFlavoredKW}, when one goes across
 the $\rho=0$ point, the amplitude of the string coupling is
 continuous, but its phase jumps by $(2n+1)\pi$, $n\in\mathbb{Z}$.

 Now let us consider the other equations.
 Computing the difference between the
 first and second equations on the left hand side of
 eq\eqref{BCCNReq+sol}, and let $u=2f-2g$,
 we will get
 \beq{}
 \dot{u}=6-6e^u+\frac{1}{\rho+\nwsgn{\rho}\cdot c}
 \label{flavoredKWueq}
 \eeq
 This equation has solution
 \beq{}
 e^u=\frac{6(\rho+\nwsgn{\rho}\cdot c)}
 {6(\rho+\sgn{\rho}\cdot c)-1+(1+\nwsgn{\rho}\cdot C_u)e^{-6\rho}}
 \label{e2u}
 \eeq
 where $C_u$ is a second integration constant whose combination
 with $c$, i.e. $6c/(6c+C_u)$ determines
 the relative size of the $U(1)$ bundle to that of the base
 space of $S^2\times S^2$ of the conifold $\hat{T}^{1,1}$ at the
 $\rho=0$ point.
 This constant plays the same role as the parameter $c_1$
 of reference \cite{BCCNR0612}.

 \begin{figure}[h]
 \hspace{-2mm}\includegraphics[scale=1]{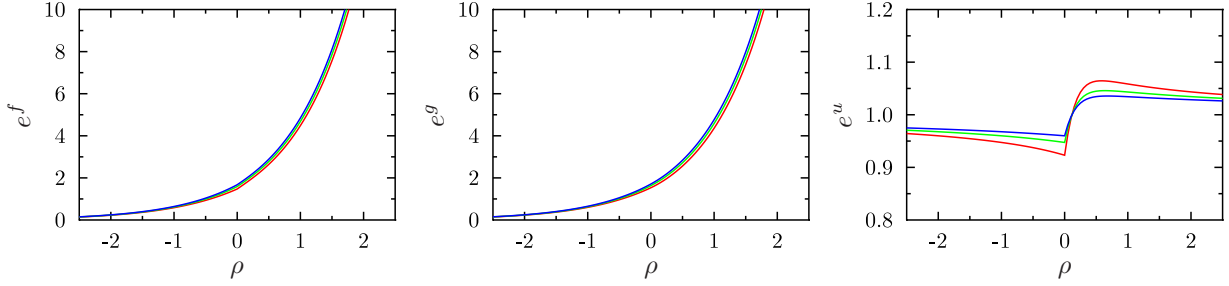}
 \caption{Metric functions of the extended BCCNR-KW solution.
 $C_u=1$, this corresponds to the $c_1=0$ case of
 \cite{BCCNR0612}. red line has $c=2$, green line $c=3$ and the blue line $c=4$}
 \label{FigureMetricF}
 \end{figure}
 Using the above $e^u=e^{2f-2g}$'s expression \eqref{e2u}, after some necessary
 algebras, we can get from the first order equations
 \eqref{BCCNReq+sol} all the remaining functions of
 the metric ansatz \eqref{BCCNRmetricAnsatz}. The result is
 summarized in the following,
 \beq{}
 \left\{
 \begin{array}{l}
 e^\phi=-\frac{4\pi}{3N_f}\frac{1}{\rho\pm c},
 \\
 e^g=\big|\left[6(\rho\pm c)-1\right]e^{6\rho}+(1\pm
 C_u)\big|^{\frac{1}{6}},
 \\
 e^f=e^{3\rho+\frac{1}{2}\ln|6(\rho\pm c)|}e^{-2g},
 \\
 h(\rho)=27\pi N_c\int_\rho^\infty d\rho e^{-4g}+C_h,
 \\
 r(\rho)=\int_{-\infty}^\rho d\rho e^f
 ,\ h_0=27\pi N_c\int_0^\infty d\rho e^{-4g}
 \end{array}
 \right.
 \label{extendedKWBCCNRSol}
 \eeq
 where $C_h$ is our third integration constant which determines the
 value of $h$ when $\rho\rightarrow\infty$.
 The sign choice in eqs\eqref{extendedKWBCCNRSol} is, when $\rho<0$, the ``$-$'' is chosen, otherwise the ``$+$''
 sign is chosen. Figure \ref{FigureMetricF} displays the behavior
 of the metric functions $e^g$, $e^f$ and their ration square
 $e^u\equiv e^{2f-2g}$. Obviously, this three functions are
 continuous but not smooth --- at the $\rho=0$ point. This
 non-smoothness will lead to the discontinuousness of the
 Ricci scalar and its higher powers.

 In the $\rho<0$ region, setting $c=0$, the functions in
 eqs\eqref{extendedKWBCCNRSol} will give the
 same metric as reference \cite{BCCNR0612} with its $c_1$ set
 as $c_1=C_u-1$, which has been quoted in the right
 side of eq\eqref{BCCNReq+sol}.
 Just as we will point out in the next section,
 as long as $c\neq0$, eq\eqref{extendedKWBCCNRSol} is regular
 at $\rho=0$ (reference \cite{BCCNR0612} did the contrary so its
 Ricci scalar diverges in the way $(-\rho)^{-\frac{5}{2}}$
 as $\rho\rightarrow0^-$ --- in the string frame.
 By the statement there, $\rho\rightarrow0^-$
 is a Landau pole of the dual gauge theory in the ultraviolet region).
 So it is the specific choice of the dilaton
 field's integration constant that leads to the singularity of
 reference \cite{BCCNR0612} at $\rho=0$. In our solutions, we removed
 this singularity by choosing a nonzero dilaton field integration
 constant. And by allowing the string coupling constant can
 take negative values, we extend BCCNR's solution
 beyond the critical point $\rho=0$. Nevertheless,
 in the extended solutions, $\rho=0$ is a discontinuous point.
 Assuming that the negative coupling constants are just as physical as
 the positive ones, then the two sides of this point is connected by a
 phase transition. Although the metric is continuous, since
 the discontinuity of the dilaton field,
 the super-gravity background as a whole is discontinuous, so
 this transition is a first order one.
 One can also imagine the existence of physical mechanisms which can
 transmute the phase of negative coupling constants into the
 the redefinition of the corresponding super-gravity axion/gauge
 theory $\theta$ angles, but we are not sure the existence of such mechanisms.

 \subsection{Asymptotic Behaviors of the extended Solution}

 To obtain the solution \eqref{extendedKWBCCNRSol}, we do not
 require the validity of super-gravity descriptions of
 the dual gauge theory in the far IR and deep UV region.
 What we required is, the validity of super-gravity
 descriptions of the dual theory from IR to UV,
 especially when one goes across the $\rho=0$ point.
 Nevertheless, getting known the asymptotical behavior
 of the super-gravity background in the far IR and deep UV
 region as well as on the intermediate energy scale is
 meaningful.

 When $\rho\rightarrow\infty$, eqs\eqref{extendedKWBCCNRSol} give
 a very interesting geometry with a complex valued, running dilaton,
 \beq{}
 \rho\rightarrow\infty:
 &&r\rightarrow e^{\rho+c},
 \nwl\\
 &&h\rightarrow\frac{27\pi N_c}{4r^4}+C_h
 ,\ e^{2g}\rightarrow r^2,\ e^{2f}\rightarrow r^2,
 \nwl\\
 &&e^\phi\rightarrow\frac{-4\pi}{3N_f\ln r},\ \mathrm{or}
 \nwl\\
 &&\phi=\ln\left[\frac{4\pi}{3N_f\ln r}\right]+i(2n+1)\pi, n\in\emph{Z}
 \label{ubehaviorBCCNRgeometry1}
 \eeq
 So, in the $\rho\rightarrow\infty$ limit, the
 geometry given by functions in eqs\eqref{extendedKWBCCNRSol}
 is almost the same as that formed by $N_c$ D3-branes placed
 at the tip of a conifold, except a complex-valued dilaton field.
 Since the imaginary part of the dilaton field is
 constant, we wish it can be absorbed into the redefinition of the
 corresponding axion field.
 When this is done, a positive string coupling $e^\phi$
 will be obtained. Physically, the
 asymptotical behavior \eqref{ubehaviorBCCNRgeometry1} implies
 that, when the flavor
 branes are placed at the $r=0$ point,
 their effects on the resulting system's property is of both
 local and topological. Locally, they lead to the running
 of the string coupling constant's amplitude. Topologically, they change
 the sign of the string coupling constant, or equivalently, they
 give the dilaton field an imaginary part of amount $i(2n+1)\pi$.

 To see the asymptotical behavior of the solution
 \eqref{extendedKWBCCNRSol} in the far IR region,
 we need to distinguish the $C_u=1$ (equivalently
 ref\cite{BCCNR0612}'s parameter $c_1=0$) case from other cases
 specifically, because in this case
 \beq{}
 \rho\rightarrow-\infty:
 &&r\rightarrow\big[|6\rho|e^{6\rho}\big]^{\frac{1}{6}}\rightarrow0,
 \nwl\\
 &&e^{2g}\rightarrow\big[|6\rho|e^{6\rho}\big]^{\frac{1}{3}}\rightarrow r^2,
 \ e^{2f}\rightarrow\big[|6\rho|e^{6\rho}\big]^{\frac{1}{3}}\rightarrow r^2
 \nwl\\
 &&h\rightarrow\frac{27\pi
 N_c}{4\big[|6\rho|e^{6\rho}\big]^{\frac{2}{3}}}\rightarrow\frac{27\pi N_c}{4r^4}
 ,\ e^\phi\rightarrow 0
 \eeq
 While in the $C_u\neq1$ case
 \beq{}
 \rho\rightarrow-\infty:
 &&r\rightarrow0,
 \nwl\\
 &&e^{2g}\rightarrow(1-C_u)^{\frac{1}{3}},
 \ e^{2f}\rightarrow
 e^{6\rho+\ln|\rho-c|}(1-C_u)^{-\frac{2}{3}}\rightarrow0,
 \nwl\\
 &&h\rightarrow(1-C_u)^{-\frac{2}{3}}(-\rho)
 ,\ e^\phi\rightarrow 0
 \eeq
 Obviously, in this limit, the quantity $e^{g-f}$ diverges. To
 analyze the singular/regular properties of the solution, we need to
 know the expressions of the Ricci scalar explicitly.
 Using BPS equations on the left side of eq\eqref{BCCNReq+sol},
 the Ricci scalar of the metric \eqref{BCCNRmetricAnsatz} can be
 calculated to be\footnote{we use the curvature convention
 $R^\sigma_{\rho\mu\nu}$ $=$ $\Gamma^\sigma_{\rho\mu,\nu}-\Gamma^\sigma_{\rho\nu,\mu}+\cdots$}
 \beq{}
 R=h^{-\frac{1}{2}}e^{-2g}\left[
 24-4e^{2f-2g}-4\left(\frac{3N_f}{4\pi}\right)e^{\phi}-\left(\frac{3N_f}{4\pi}\right)^2e^{2\phi-2f+2g}
 \right]
 \label{fullRhicciScalarBCCNR1}
 \eeq
 So, as $\rho\rightarrow-\infty$, i.e. $r\rightarrow0$,
 if (i) $C_u=1$, then the geometry is regular and asymptotically
 approaches $AdS_5\times T^{1,1}$; (ii) $C_u\neq1$, then the
 relative size between the $U(1)$ factor and the $S^2\times S^2$
 part of $\hat{T}_{1,1}$ --- which has the topoloty of $U(1)$ bundles
 over base $S^2\times S^2$ --- is infinitely squashed.
 In this case, $\rho=-\infty$, i.e. $r=0$ is a
 singular point of the
 flavored Klebanov-Witten geometry. By the statement of \cite{BCCNR0612}, this
 is a ``good'' IR singularity of the dual field theory. Our
 analysis here support this statement in the following way,
 this singularity can be avoided by a specific integration
 constant choice, i.e. $C_u=1$. So it can be looked as an
 artificial singularity instead of an essential one.

 {\bf Remarks:} Comparing eq\eqref{fullRhicciScalarBCCNR1}
 with the eq(2.42) of reference \cite{BCCNR0612}, one may
 wonder the difference between the two. This difference
 originate from two facts. The first is that the Ricci scalar
 provided by \cite{BCCNR0612} is in the string frame while ours is in
 the Einstein frame. The second is that \cite{BCCNR0612} provides
 only parts of the Ricci scalar which come from contributions of the
 flavor branes. For comparison we also computed the full Ricci
 scalar of metric \eqref{BCCNRmetricAnsatz} in the string frame, the result is
 \beq{}
 R^{s}=2h^{-\frac{1}{2}}e^{-2g-\frac{1}{2}\phi}
 \left[2(6-e^{2f-2g})+7\frac{3N_f}{4\pi}e^{\phi-2g}
 +4\left(\frac{3N_f}{4\pi}\right)^2e^{2\phi-2f+2g}\right]
 \eeq
 Either from this expression, or from
 eq\eqref{fullRhicciScalarBCCNR1}, we know that
 Ricci scalars of the flavored geometry \eqref{extendedKWBCCNRSol}
 can be looked as a second order polynomial of $N_fe^\phi$.

 Finally, we point out that, the solution \eqref{extendedKWBCCNRSol}
 is regular at $\rho=0$ point as long as $c\neq0$. Note
 \beq{}
 \rho\rightarrow0:
 &&e^{-2g}\rightarrow(6c+C_u)^{-\frac{1}{3}}
 \nwl\\
 &&e^{u}\equiv e^{2f-2g}\rightarrow\frac{6c}{6c+C_u},\ e^\phi\rightarrow\pm\frac{1}{c}
 \nwl\\
 &&R\rightarrow \frac{(h_0+C_h)^{-\frac{1}{2}}}{(6c+C_u)^{\frac{1}{3}}}
 \left(24-\frac{24c}{6c+C_u}-\frac{4}{\pm c}+\frac{1}{c^2}\frac{(6c+C_u)^2}{36c^2}
 \right)
 \eeq
 we see that when $\rho\rightarrow0$, the Ricci scalar
 is non-singular at $\rho=0$. In the above equations, when $\pm$
 sign appears, ``$+$'' sign corresponds to the limit of $R$ when
 $\rho\rightarrow0^+$, while ``$-$'' sign corresponds to the
 limit of $R$ when $\rho\rightarrow0^+$. Obviously, at the point
 $\rho=0$, the Ricci scalar is discontinuous.

 \section{Extension of BCCNR flavored KS geometry}

 \subsection{One comment about the BCCNR-KS solution itself}
 In their second work \cite{BCCNR0706}, BCCNR considered the
 flavoring of the KT/KS geometries. We focus here on their
 flavoring of the KS geometry. By similar smearing procedure
 as that for the flavoring of KW geometry, BCCNR set up ansatz for
 the metric and various form fields of the flavored KS geometry as follows,
 \beq{}
 &&\hspace{-3mm}ds^2=h^{-\frac{1}{2}}dx_{1,3}^2+h^{\frac{1}{2}}(\frac{1}{9}e^{2G_3}d\tau^2+e^{2G_3}d\tilde{T}_{1,1}^2)
 \nwl\\
 &&\hspace{-3mm}d\tilde{T}_{1,1}^2=e^{2G_1-2G_3}(\sigma_1^2+\sigma_2^2)
 \nwl\\
 &&\hspace{10mm}+e^{2G_2-2G_3}\left[(\omega_1+g\sigma_1)^2+(\omega_2+g\sigma_2)^2\right]+\frac{1}{9}(\omega_3+\sigma_3)^2
 \nwl\\
 &&\hspace{-3mm}\sigma_1=d\theta_1,\
 \sigma_2=\sin\theta_1d\varphi_1,\ \sigma_3=\cos\theta_1d\varphi_1
 \nwl\\
 &&\hspace{-3mm}\omega_1=\sin\psi\sin\theta_2d\varphi_2+\cos\psi d\theta_2,
 \nwl\\
 &&\hspace{-3mm}\omega_2=-\cos\psi\sin\theta_2d\varphi_2+\sin\psi d\theta_2,
 \ \omega_3=d\psi+\cos\theta_2d\varphi_2
 \label{flavoredKSansatz}
 \eeq
 \beq{}
 &&\hspace{-3mm}F_1=\frac{N_f}{4\pi}g^5,
 \nwl\\
 &&\hspace{-3mm}H_3=dB_2,\ B_2=\frac{M}{2}\left[fg^1\wedge
 g^2+kg^3\wedge g^4\right],
 \nwl\\
 &&\hspace{-3mm}F_3=\frac{M}{2}\bigg\{
 F^\prime dr\wedge(g^1\wedge g^2+g^3\wedge g^4)
 \nwl\\
 &&\hspace{5mm}+g^5\wedge\left[(F+\frac{N_f}{4\pi}f)g^1\wedge g^2+(1-F+\frac{N_f}{4\pi}k)g^3\wedge g^4\right]
 \bigg\}
 \nwl\\
 &&\hspace{-3mm}F_5=dh^{-1}(r)\wedge dx^0\wedge...\wedge dx^3
 +\mathrm{Hodge\ dual}
 \label{flavoredKSFormansatz}
 \eeq
 where $g^{1,2,3,4,5}$ are constant combinations of the $\sigma_i$ and
 $\omega_i$s involved in the metric ansatz. For details, pleas see
 reference
 \cite{BCCNR0706}. By super-symmetry analysis, BCCNR
 get first order equations and the corresponding
 solution for the various functions involved in the
 above ansatz\footnote{the over dot denotes $\frac{d}{d\tau}\equiv
 \frac{e^{G_3}}{3}\frac{d}{dr}$.},
 \beq{}
 \left\{
 \begin{array}{l}
 \dot{\phi}=\frac{N_f}{4\pi}e^\phi,
 \\
 g(g^2-1+e^{G_1-G_2})=0
 \\
 \dot{G}_1=\frac{1}{18}e^{2G_3-G_1-G_2}+\frac{1}{2}e^{G_2-G_1}-\frac{1}{2}e^{G_1-G_2}
 \\
 \dot{G}_2=\frac{1}{18}e^{2G_3-G_1-G_2}-\frac{1}{2}e^{G_2-G_1}+\frac{1}{2}e^{G_1-G_2}
 \\
 \dot{G}_3=-\frac{1}{9}e^{2G_3-G_1-G_2}+e^{G_2-G_1}-\frac{N_f}{8\pi}e^\phi
 \end{array}
 \right.,
 \left\{
 \begin{array}{l}
 e^\phi=-\frac{4\pi}{N_f}\frac{1}{\tau-\tau_0}
 \\
 e^{2G_1}=\frac{1}{4}\mu^{\frac{4}{3}}\frac{\sinh^2\tau}{\cosh\tau}\Lambda
(\tau),
 \\
 e^{2G_2}=\frac{1}{4}\mu^{\frac{4}{3}}\cosh\tau\Lambda(\tau)
 \\
 e^{2G_3}=6\mu^{\frac{4}{3}}\frac{\tau_0-\tau}{[\Lambda(\tau)]^2}
 \\
 g=\frac{1}{\cosh\tau},\ 0<\tau<\tau_0
 \end{array}
 \right.
 \eeq
 \beq{}
 \left\{
 \begin{array}{l}
 \dot{k}=e^\phi\left[F+\frac{N_f}{4\pi}f\right]\coth^2\frac{\tau}{2}
 \\
 \dot{f}=e^\phi\left[1-F+\frac{N_f}{4\pi}k\right]\tanh^2\frac{\tau}{2}
 \\
 \dot{F}=\frac{1}{2}e^{-\phi}(k-f)
 \end{array}
 \right.,
 \hspace{10mm}\left\{
 \begin{array}{l}
 e^{-\phi}k=\frac{\tau\coth\tau-1}{\sinh\tau}(\cosh\tau+1)
 \\
 e^{-\phi}f=\frac{\tau\coth\tau-1}{\sinh\tau}(\cosh\tau-1)
 \\
 F=\frac{\sinh\tau-\tau}{2\sinh\tau}.
 \end{array}
 \right.\label{BCCNRformFieldsKS}
 \eeq

 \beq{}
 &&\hspace{-3mm}\dot{h}e^{2G_1+2G_2}=
 -\frac{M^2}{4}\left[f+(k-f)F+\frac{N_f}{4\pi}kf\right]+\mathrm{const}.
 ,\nwl\\ &&\hspace{-3mm}h=\frac{M^2}{4}\int^\tau d\tau e^{-2G_1-2G_2}
 \left[f+(k-f)F+\frac{N_f}{4\pi}kf\right]
 \label{warpFactorBCCNRKS}
 \eeq

 Our comment on BCCNR flavored KS geometry itself
 focuses on its form fields solution.
 We suspect the solutions
 \eqref{BCCNRformFieldsKS} may not reflect the flavor branes
 effect completely. The reason is as follows:
 when the equations satisfied by the form fields are translated
 into the following form,
 \beq{}
 &&\hspace{-3mm}(e^{-\phi}k)^{\cdot}=F[\coth(\frac{\tau}{2})]^2
 -\frac{N_f}{4\pi}\left(k-f\cdot[\coth(\frac{\tau}{2})]^2\right)
 \nwl\\
 &&\hspace{-3mm}(e^{-\phi}f)^{\cdot}=(1-F)[\tanh(\frac{\tau}{2})]^2
 -\frac{N_f}{4\pi}\left(f-k\cdot[\tanh(\frac{\tau}{2})]^2\right)
 \nwl\\
 &&\hspace{-3mm}\dot{F}=\frac{1}{2}(e^{-\phi}k-e^{-\phi}f).
 \label{e2phikfEq}
 \eeq
 we easily see that, only if
 \beq{}
 k-f\cdot[\coth(\frac{\tau}{2})]^2=0,
 \label{kfconstraint}
 \eeq
 will the equations satisfied by functions
 $e^{-\phi}k$, $e^{-\phi}f$ and $F$ reduce to the
 same form as those satisfied by functions
 $k$, $f$ and $F$ of KS geometry,
 please see eq(5.29) of \cite{KS0007}; BCCNR's
 function $e^{-\phi}k$, $e^{-\phi}f$ and $F$ have
 the same form as KS' $k$, $f$ and $F$, so they
 satisfy eq\eqref{kfconstraint} implicitly. Our question
 is, why did BCCNR imposed the constraint \eqref{kfconstraint}
 on the form fields function? or equivalently, why is
 just the solution satisfying the constraint \eqref{kfconstraint} being chosen?

 Obviously, the answer to this kind of question should be determined by
 the boundary conditions. We know that in the KS theory,
 to obtain the form fields solution, the following boundary
 conditions\footnote{We use subscript $KS$ to denote quantities in
 the Klebanov Strassler's solution. When necessary, we will use
 $BCCNR$ to denote quantities from the BCCNR solutions.}
 \beq{}
 \begin{array}{l}
 \tau\rightarrow0:
 \\ \\ \\
 \end{array}
 \begin{array}{l}
 k_{KS}\rightarrow\tau
 \\
 f_{KS}\rightarrow\tau^3
 \\
 F_{KS}\rightarrow\tau^2
 \end{array}
 \begin{array}{l}
 \\ \\;
 \end{array}
 \begin{array}{l}
 \tau\rightarrow\infty:
 \\ \\ \\
 \end{array}
 \begin{array}{l}
 k_{KS}\rightarrow\frac{1}{2}\tau
 \\
 f_{KS}\rightarrow\frac{1}{2}\tau
 \\
 F_{KS}\rightarrow\frac{1}{2}\,\,
 \end{array}\label{BCCNRformFieldsSolLimit2}
 \eeq
 especially those at
 the $\tau\rightarrow\infty$ point, are imposed on the equations of motion,
 please see eqs(5.29-5.32) of \cite{KS0007}. In BCCNR's
 framework, we cannot take the $\tau\rightarrow\infty$
 point as a sensible boundary point, because in this framework
 the $\tau$ coordinate can only take values in
 the finite range $0<\tau<\tau_0$. So a crucial question
 is, what kinds of boundary conditions, when imposed on
 eqs\eqref{e2phikfEq}, are reasonable, i.e., completely account
 for the effects of flavor branes back-reaction?

 Although we cannot answer this question definitely.
 We can consider the possibilities more general than
 BCCNR's solution. For example, discarding the constraint
 eq\eqref{kfconstraint}. In this case, we can
 write the solution of \eqref{BCCNRformFieldsKS} as
 \begin{subequations}
 \beq{}
 &&\hspace{-3mm}e^{-\phi}k=k_{KS}+p(\tau)=\left[\frac{\tau\coth\tau-1}{\sinh\tau}(\cosh\tau+1)+p(\tau)\right],
 \\
 &&\hspace{-3mm}e^{-\phi}f=f_{KS}+q(\tau)=\left[\frac{\tau\coth\tau-1}{\sinh\tau}(\cosh\tau-1)+q(\tau)\right]
 \\
 &&\hspace{-3mm}F=F_{KS}+Q(\tau)=\frac{\sinh\tau-\tau}{2\sinh\tau}+Q(\tau).
 \eeq
 \label{formFieldsSolFlavoredKS}
 \end{subequations}
 where $p$, $q$ and  $Q$ are newly introduced functions
 to measure the deviation of form fields when discarding
 the constraint \eqref{kfconstraint}. They satisfy
 \beq{}
 &&\hspace{-3mm}\dot{p}=Q\coth^2\left[\frac{\tau}{2}\right]
 -\frac{N_f}{4\pi}e^\phi\left(p-q\coth^2\left[\frac{\tau}{2}\right]\right)
 \nwl\\
 &&\hspace{-3mm}\dot{q}=-Q\tanh^2\left[\frac{\tau}{2}\right]
 -\frac{N_f}{4\pi}e^\phi\left(q-p\tanh^2\left[\frac{\tau}{2}\right]\right)
 \nwl\\
 &&\hspace{-3mm}\dot{Q}=\frac{1}{2}(p-q).
 \label{pqQequations}
 \eeq
 Obviously, if the quantity
 $\Delta\equiv p-q\coth^2\left[\frac{\tau}{2}\right]\neq0$, then this quantity
 will affects the evolution of $p$ and $q$ in the combination
 $Ne^\phi\Delta$ and eventually make functions $e^{-\phi}k$, $e^{-\phi}f$
 and $F$ to depend $\Delta$ in the following way
 \beq{}
 e^{-\phi}k=k_{KS}+c_1N_fe^\phi\Delta+c_2(N_fe^\phi\Delta)^2+\cdots
 \nwl\\
 e^{-\phi}f=f_{KS}+d_1N_fe^\phi\Delta+d_2(N_fe^\phi\Delta)^2+\cdots
 \nwl\\
 F=F_{KS}+e_1N_fe^\phi\Delta+e_2(N_fe^\phi\Delta)^2+\cdots
 \eeq
 where $c_i$, $d_i$ and $e_i$ are the relevant expansion
 coefficients.

\begin{figure}[h]
 \includegraphics[]{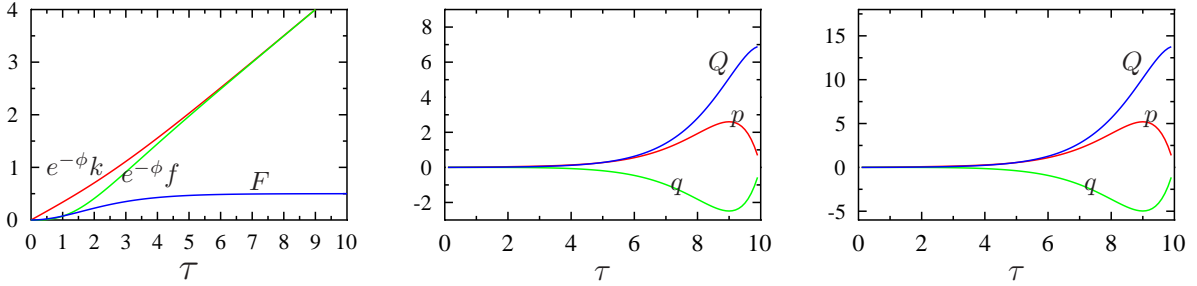}
 \caption{The left figure displays BCCNR's
 form fields solution. The middle one is the numerical
 solution of eqs\eqref{pqQequations}, with boundary conditions
 $\{p,q,Q\}_{\tau\rightarrow0}=\epsilon*\{k_{KS}, f_{KS}, F_{KS}\}_{\tau\rightarrow0}$, $\epsilon=0.05$.
 The right one comes with boundary conditions
 $\{p,q,Q\}_{\tau\rightarrow0}=\epsilon*\{k_{KS}, f_{KS}, F_{KS}\}_{\tau\rightarrow0}$, $\epsilon=0.1$.
 In these plots, we take $e^\phi=-4\pi/N_f(\tau_0-\tau)$ and $\tau_0=10$.}
 \label{FigurepqQfunctions}
 \end{figure}

 What's frustrated to us is, we cannot integrate eqs\eqref{pqQequations}
 analytically. We can only integrate it numerically. We display in
 Figure \ref{FigurepqQfunctions} the behavior of $p$, $q$ and $Q$
 which follows from a kind of special boundary conditions. From
 this figure, we can see two facts. This first is, in the small
 $\tau$ region, $p$, $q$ and $Q$ can be neglected, but in the large
 $\tau$ region, their contributions to the complete solution is
 remarkable. The second is, the magnitude of $p$, $q$, $Q$
 proportionally depend on the their values at the boundary point
 $\tau\rightarrow0$. This fact can also be accounted by the fact
 that eqs\eqref{pqQequations} is linear homogeneous ordinary differential
 equations.
 So the numerical solution tells us, if on any point, including the
 boundary point $\tau=0$ and $\tau=\tau_0$, the constraint
 \eqref{kfconstraint} is not obeyed\footnote{Although our boundary
 condition satisfy the constraint \eqref{kfconstraint}, since
 $p$, $q$ and $Q$ satisfy different equations as $k_{ks}$, $f_{ks}$
 and $F_{ks}$, they will make the complete $e^{-\phi}k$, $e^{-\phi}f$
 and $F$ do not satisfy eqs\eqref{kfconstraint} in the internal
 point.}, they will not be obeyed
 in the internal region of $0<\tau<\tau_0$. But if we are
 interested in the small $\tau$ region, then regardless of the
 boundary, we can always take the eq\eqref{kfconstraint} as
 a valid constraint.

\begin{figure}[h]
 \begin{minipage}[c]{0.62\textwidth}
 \caption{The effective D3-brane charges as functions
 of the holographic radial coordinate. The red line
 is BCCNR's result, the green and blue lines are our
 result when discarding the constraint eq\eqref{kfconstraint}.
 For comparisons, we also display the effective D3-brane charges
 of the unflavored KS theory in black line. The parameter
 choice of this figure is the same as that of
 Figure \ref{FigurepqQfunctions}.}
 \label{FigureNeff}
 \end{minipage}
 \begin{minipage}[c]{0.35\textwidth}
 \includegraphics[scale=1.1]{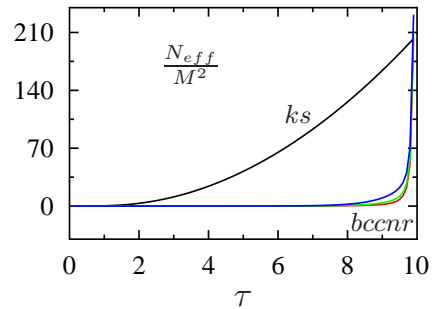}
 \end{minipage}
 \end{figure}

 From reference \cite{KS0007}, we know that, the duality
 cascading can be ascribed to the varying of effective D3-brane
 charges with respect to the holographic radial coordinate.
 By the effective D3-brane charge definition of \cite{BCCNR0706},
 \beq{}
 N_{eff}(\tau)=N_0+\frac{M^2}{\pi}\left[
 f+(k-f)F+\frac{N_f}{4\pi}kf\right]
 \label{effD3charge}
 \eeq
 we know that, if the functions $k$, $f$ and
 $F$ deviate from BCCNR's results, then the corresponding
 duality cascading should also be different from that of BCCNR.
 We displayed in Figure \ref{FigureNeff} the dependence of the
 effective D3-brane charges on the radial coordinate. From the
 figure we see that, although the deviation of functions $k$, $f$
 and $F$ deviate from BCCNR's result remarkably in the large $\tau$
 region, the corresponding $N_{eff}$ does not do so. As the
 conclusion, we say that discarding the constraint eq\eqref{kfconstraint}
 does not change the dual cascading described by reference
 \cite{BCCNR0706} qualitatively.
 \hspace{-3mm}

 \subsection{Extension of BCCNR-KS solution  --- beyond the duality wall}

 This section we study the extension of BCCNR flavored
 KS solution into negative coupling constant region.

 First, we note that in the KS solution, the sensible
 region of the radial coordinate is $0<\tau<\infty$\footnote{
 Only from the metric functions, we cannot look this out. This
 fact is necessary to assure the positivity of the form fields
 in the KS theory.}. While
 in BCCNR's solution, to assure the reality of
 the dilaton field, the sensible value of $\tau$ coordinate is
 $0<\tau<\tau_0$. If we allow the string coupling can take negative values,
 or equivalently the dilaton field can carry an imaginary part of $\imath(2n+1)\pi$,
 then we can extend BCCNR's solution into the $0<\tau<\infty$
 region:
 \beq{}
 e^\phi=-\frac{4\pi}{N_f}\frac{1}{\tau-\tau_0\pm c}
 ,\ 0<\tau<\infty
 \label{dilatonFlavorKS}
 \eeq
 The sign choice here is similar to that in eq\eqref{extendedKWBCCNRSol},
 i.e., when the ``$\pm$'' sign appears, the ``$-$'' sign is
 for the $0<\tau<\tau_0$ part of the relevant expressions while
 the ``$+$'' sign is for the $\tau_0<\tau<\infty$ part. Also similar
 to the requirement there, we require $c>>1$ and
 $N_ce^\phi|_{\tau=\tau_0}\approx \frac{4\pi N_c}{N_fc}>>1$ simultaneously so
 that super-gravity descriptions at the $\tau=\tau_0$ point is valid.
 Obviously, when one goes across
 the $\tau=\tau_0$ point, the amplitude of the string coupling is
 continuous, but its phase jumps by $(2n+1)\pi$, $n\in\mathbb{Z}$.
 We assume that, in the negative coupling constant region
 $\tau_0<\tau$, as long as $|e^\phi|<<1$ and $|N_ce^\phi|>>1$,
 then super-gravity descriptions are still valid.
 \begin{figure}[h]
 \begin{center}\includegraphics[]{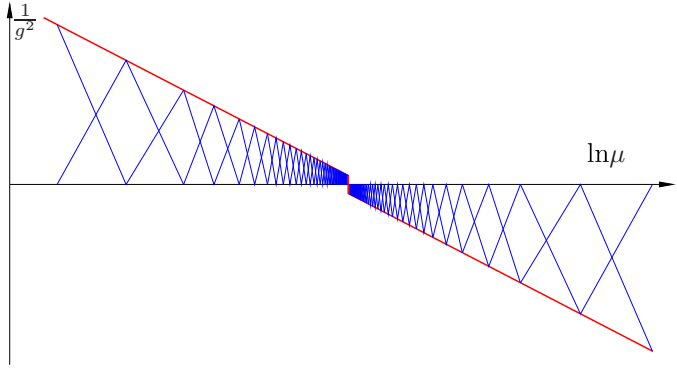}\end{center}
 \caption{
 The running gauge coupling constants as functions of the logarithm
 of energy scales in the extended BCCNR flavored cascading
 gauge theory. The blue lines are the inverse of the squared gauge
 coupling constants $\frac{1}{g_{i}^2}$ while the red line is their sum,
 $i=l,s$ characterize
 coupling constants corresponding to the large and small rank gauge groups.}
 \label{figureDualityWall}
 \end{figure}

 On the dual gauge theory side, the negative string coupling
 constant means that the gauge coupling constants are also
 negative. Or equivalently, the gauge coupling constants carry
 imaginary part of $i(2n+1)\pi$. Just as we expressed in the previous
 sections, we are not sure the existence of
 physical mechanism which can transmute this imaginary part into
 the gauge theory $\theta$ parameter, and make the negative
 coupling constants sensible concepts. We just simply assume
 that the negative coupling constants are just as physical
 as positive ones, and both the super-gravity descriptions and
 the AdS/CFT correspondence are valid. Under this assumption, BCCNR's
 duality cascading can be extended beyond the duality wall
 marked by the radial coordinate $\tau_0$. We displayed in Figure
 \ref{figureDualityWall} the extended duality cascading picture.
 From this figure we see that, below the duality wall, the slope of
 the $\frac{1}{g_i^2}$ versus energy scale becomes larger and larger
 as RG flows up. However, as long as it goes beyond the duality
 all, the trend reverses.

 Second, we consider the extension of BCCNR's metric
 functions $G_1$, $G_3$, $G_3$. Construct the following combinations,
 \beq{}
 &&\hspace{-3mm}w\equiv G_1-G_2:\ \dot{w}=e^{-w}-e^w,
 \nwl\\
 &&\hspace{-3mm}v\equiv G_1+G_2+G_3:\ \dot{v}=e^{-w}-\frac{N_f}{8\pi}e^\phi,
 \nwl\\
 &&\hspace{-3mm}u\equiv 2G_3-G_1-G_2:\ \dot{u}=-\frac{1}{3}e^u+2e^{-w}-\frac{N_f}{4\pi}e^\phi.
 \label{uvwDefinitionEq}
 \eeq
 where $w$ describes the relative size of the two $S^2$ factors in the
 base space of a deformed conifold which has the topology
 of a $U(1)$ fibre over base $S^2\times S^2$; $v$ is related to the total
 \emph{volume} of the conifold; while $u$ corresponds to the squashing
 of the $U(1)$ fibre of the base relative to the $S^2\times S^2$ factor.
 Integration of these equations
 should distinguish three different cases. The first is
 when $G_1\equiv G_2$, this case will give metric functions of the
 KT solution, which is equivalent to those of the previous section.
 The second case is when
 $G_1<G_2$, the last one is when $G_1>G_2$. Since the two factor
 $S^2$s of the conifold's base is symmetric, the metric functions
 in the case of $G_1>G_2$ can be obtained from those of the
 $G_1<G_2$ by simply interchanging the position of the two $S^2$
 factors. Considering this fact, we will assume that $G_1<G_2$ in the
 following. Under this assumption, the solution
 to eq\eqref{uvwDefinitionEq} can be written as
 \begin{subequations}
 \beq{}
 &&\hspace{-3mm}e^w
 =\frac{\sinh(\tau+c_w)}{\cosh(\tau+c_w)}\label{wsolFlavoredKS}
 ,\\
 &&\hspace{-3mm}e^v
 =c_v\sinh(\tau+c_w)|(\tau-\tau_0\pm c)|^{\frac{1}{2}}
 \label{usolFlavoredKS}
 ,\\
 &&\hspace{-3mm}e^u
 =\frac{24(\tau-\tau_0\pm c)\sinh^2(\tau+c_w)}
 {2(\tau-\tau_0\pm c)\sinh2(\tau+c_w)-\cosh2(\tau+c_w)
 -2(\tau+c_w)(\tau-\tau_0\pm c)+c_u}
 \label{vsolFlavoredKS}
 \eeq\label{uvwSol}
 \end{subequations}
 where $c_w$, $c_v$ and $c_u$ are integration constants. $c_w$
 specifies the relative size between the two $S^2$ factors in the
 conifold's base space at $\tau=0$ point.
 $c_v$ is related to the total volume of the conifold's base at $\tau=0$.
 It plays the same role as the parameter $\mu$ does in the eqs(3.7)
 of reference \cite{BCCNR0706}. $c_u$ specifies the relative
 size of the $U(1)$ bundle to the $S^2\times S^2$ factor at $\tau=0$
 point. The constant $c_w$ can always be set to zero by translating
 redefinition of the $\tau$ coordinate. We will do so in the following.
 In this case, if $c_u=1$, $e^u|_{\tau=0}=0$, otherwise $e^u$ diverges at
 the $\tau=0$ point.

 \begin{figure}[h]
 \includegraphics[]{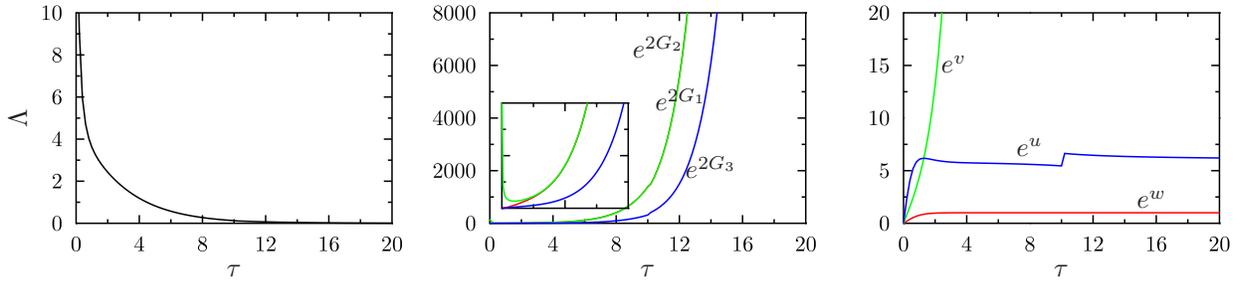}
 \caption{Numerical behavior of the metric functions $e^{G_1}$,
 $e^{G_2}$, $e^{G_3}$ and their three specific combinations.
 parameter choices are, $\tau_0=10$, $c=5$, $c_u=1$.}
 \label{FigureG123functions}
 \end{figure}
 Substituting eqs\eqref{uvwSol} into \eqref{uvwDefinitionEq}
 and setting $c_w=0$, we get manifest expressions for the metric functions
 \begin{subequations}
 \beq{}
 &&\hspace{-6mm}e^{2G_1}=e^{w+\frac{2}{3}v-\frac{1}{3}u}
 =\frac{c_v\sinh^2\tau}{\cosh\tau}
 \Lambda(\tau),
 \\
 &&\hspace{-6mm}e^{2G_2}=e^{-w+\frac{2}{3}v-\frac{1}{3}u}
 =c_v\cosh\tau\Lambda(\tau),
 \\
 &&\hspace{-6mm}e^{2G_3}=e^{\frac{2}{3}v+\frac{2}{3}u}
 =\frac{c_v|\tau-\tau_0\pm c|}{[\Lambda(\tau)]^2},
 \\
 &&\hspace{-6mm}\Lambda\equiv
 \left[\frac{2(\tau-\tau_0\pm c)\sinh2\tau-\cosh2\tau
 -2\tau(\tau-\tau_0\pm c)+c_u}
 {\mathrm{sgn}[\tau-\tau_0]\cdot\sinh^3\tau}\right]^{\frac{1}{3}}
 \eeq
 \end{subequations}
 If we set $c=0$ and
 restrict the sensible region of $\tau$ to be $0<\tau<\tau_0$,
 then the above expressions reduce to the equations (3.6-3.7) of reference
 \cite{BCCNR0706}. We display in figure \ref{FigureG123functions}
 numerical behaviors of the three metric functions and their
 ratios/products. From the figure, we see that on the transition
 point between the negative and positive coupling constant region
 $\tau=\tau_0$, the metric functions are continuous but not smooth.
 Nevertheless, since the form fields are not continuous at this
 point, see the following,
 the super-gravity background as a whole is discontinuous. So
 if the the negative coupling region is physical and it is connected
 with the positive coupling region through a phase transition, then the
 transition is a first order one.

 \begin{figure}[h]
 \begin{center}\includegraphics[]{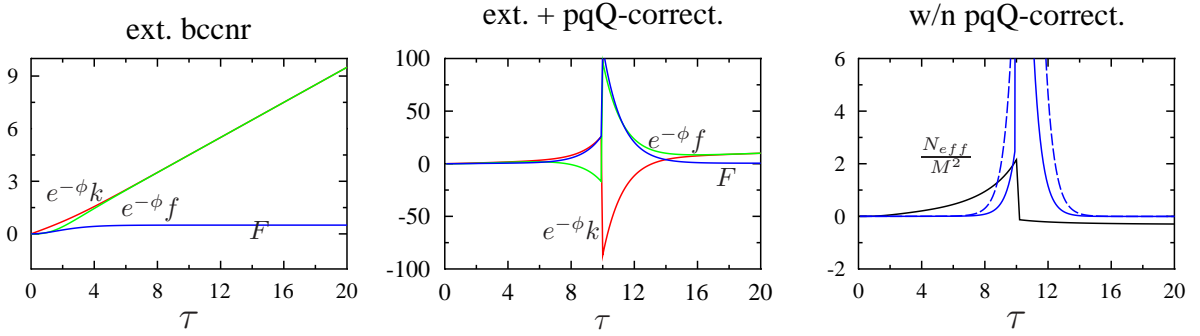}\end{center}
 \caption{
 Left, simply extending BCCNR's form fields solution
 into the negative coupling constant region $\tau_0<\tau$.
 Its corresponding effective D3-brane charges $N_{eff}$ are displayed
 in the right figure with black lines. Middle, extending the
 BCCNR's form fields solution into the negative coupling constant
 region, but allowing them to deviate from BCCNR's solution, which
 is the same as KS' unflavored solution except an $e^{-\phi}$ factor, at the boundary
 $\{p,q,Q\}_{\tau\rightarrow0}$ $\rightarrow$ $\epsilon*\{k,f,F\}_{ks,\tau\rightarrow0}$,
 $\{p,q,Q\}_{\tau\rightarrow\infty}$ $\rightarrow$
 $\epsilon*\{k,f,F\}_{ks,\tau\rightarrow\infty}$, $\epsilon=0.05$.
 The corresponding $N_{eff}$ is displayed in the
 right figure with blue lines. The dashed line in the right
 is $N_{eff}$ calculated with $\epsilon=0.1$. The
 other parameter in this figure is $N_0=0$, $\tau_0=10$, $c=5$.
 The height of blue lines, both dashed and un-dashed, in the right figure is reduced by a
 factor of $0.01$.
 }
 \label{figurekfFext}
 \end{figure}
 Finally, we turn to the extension of the flavored form fields. If
 we simply extend BCCNR's form fields solution into the negative
 coupling constant region $\tau_0<\tau$, then their expressions are
 completely the same as eqs\eqref{BCCNRformFieldsKS}, but with
 the radial coordinate taking values in the $0<\tau<\infty$ region.
 This kind of form fields will make the effective D3-brane charges
 ,defined by eq\eqref{effD3charge}, in the $\tau_0<\tau<\infty$ be
 negative, please see Figure \ref{figurekfFext}. However, if we
 allow the flavored form fields ($e^{-\phi}k$, $e^{-\phi}f$ and $F$) can
 be different from the unflavored ones ($k_{ks}$, $f_{ks}$ and $F$),
 see eqs\eqref{formFieldsSolFlavoredKS}, then the difference $p$, $q$
 and $Q$ will satisfy eqs\eqref{pqQequations}. Letting them
 be infinitesimal fractions of $k_{ks}$, $f_{ks}$ and $F_{ks}$
 on the boundary $\tau\rightarrow0$ and $\tau\rightarrow\infty$,
 they will become very large relative to the latter ones in
 the middle $\tau$ region. On the point $\tau=\tau_0$,
 when this $p$, $q$ and $Q$ corrections are
 included, the resulting form fields even become discontinuous
 functions of $\tau$. A very interesting facts is, when these corrections are included,
 the resulting effective D3-branes charge become positive in the
 whole $0<\tau<\infty$ region, please see Figure \ref{figurekfFext}
 \footnote{On the boundary point $\tau=0$ and $\tau=\infty$,
 the effective D3-branes charge is zero.}.

 With the above extended metric functions $e^{G_i}$, and form
 fields solution $k$, $f$ and $F$, we can get the warp factor
 $h(\tau)$ by simply substituting them
 into eq\eqref{warpFactorBCCNRKS} and make the integration.
 To this point, we finish the extension of BCCNR's flavored KS
 geometry to the negative coupling constant regions. For
 comparisons with the extended BCCNR-KW geometry and for
 future references, we calculated the ricci
 scalar of the BCCNR flavored KS geometry. The result is as follows,
 \beq{}
 &&\hspace{-3mm}R=e^{-2(G_1+G_2+G_3)}\Bigg\{\left[\frac{
    81\,e^{4\,{G_1}} - 81\,e^{4\,{G_2}} - e^{4\,{G_3}} +
         36e^{2G_2 + 2G_3}
         }{9h^{1/2}}\right.
 \nwl\\
 &&\hspace{15mm}\left.-\frac{
      9M^2\,\left( e^{-\phi}[k-f]^2 + 2e^\phi F^2\coth^2[\frac{\tau}{2}] +
         2e^{\phi }[1-F]^2\tanh^2[\frac{\tau}{2}] \right)  }{16
    h^{3/2}}\right]
 \nwl\\
 &&\hspace{17mm}-\frac{N_f}{4\pi}e^\phi\left[\frac{2e^{G_1+G_2+2G_3}}{h^{1/2}}+
  \frac{9M^2
     \left(k[1-F]\tanh^2[\frac{r}{2}]+fF\coth^2[\frac{\tau}{2}]\right) }
     {4h^{3/2}}\right]
 \nwl\\
 &&\hspace{17mm}+9\left(\frac{N_f}{4\pi}\right)^2e^{2\phi}\left[\frac{e^{2G_1+2G_2}}{h^{1/2}} -
  \frac{e^{-\phi}M^2(k^2\tanh^2[\frac{\tau}{2}]+f^2\coth^2[\frac{\tau}{2}])}{8h^{3/2}}
     \right]
 \Bigg\}
 \eeq

 \beq{}
 &&\hspace{-3mm}R^s=e^{-\frac{1}{2}\phi-2(G_1+G_2+G_3)}\Bigg\{\left[\frac{81e^{4G_1}-81e^{4G_2}-e^{4G_3}+36e^{2(G_2+G_3)}}
  {9h^{1/2}}\right.
 \nwl\\
 &&\hspace{15mm}\left.-\frac{
    9M^2(e^{-\phi}[k-f]^2+2e^{\phi}F^2\coth^2\left[\frac{r}{2}\right]+
      2e^{\phi}[1-F]^2\tanh^2\left[\frac{r}{2}\right])
      }{16h^{3/2}}\right]
 \nwl\\
 &&\hspace{17mm}+\frac{N_f}{4\pi}e^\phi\left[\frac{7e^{G_1+G_2+2G_3}}{h^{1/2}} -
  \frac{9M^2(k[1-F]\tanh^2\left[\frac{r}{2}\right]+fF\coth^2\left[\frac{r}{2}\right]) }
     {4h^{3/2}}\right]
 \nwl\\
 &&\hspace{17mm}+9\left(\frac{N_f}{4\pi}\right)^2e^{2\phi}\left[\frac{10e^{2G_1+2G_2}}{h^{1/2}} -
  \frac{e^{-\phi}M^2
     (k^2\tanh^2\left[\frac{r}{2}\right]+f^2\coth^2\left[\frac{r}{2}\right])}{8h^{3/2}}\right]
 \Bigg\}
 \eeq
 From these expressions, we see that, since the appearance of
 the fractional branes, marked by the parameter $M$, the Ricci scalars
 do not have the exact form of two order polynomials of $N_fe^\phi$.
 If we study the singular/regular behavior of $R$, we will see that it
 is regular at $\tau=\tau_0$ but not continuous; it is regular both
 at $\tau=0$ and $\tau=\infty$.

 \section{Conclusions}

 We extend the BCCNR flavored KW/KS geometries into negative
 coupling constant region. In the extended BCCNR-KW geometry: the
 positive coupling constant region preserves all the characters
 of the original theory; while the negative coupling constant region
 has the asymptotical geometry of multi D3-branes placed at the apex
 of the a conifold. The positive and negative coupling constant region
 have continuous, although not smooth, super-gravity metric but
 discontinuous coupling constant. In the extended BCCNR-KS
 geometry: the positive coupling constant region also preserves the
 main characters of the original theory, but we provide a comment
 which may be looked as a supplement to BCCNR's solution in the
 form fields sector; in the negative coupling constant region, the extended solution
 asymptotically becomes the unflavored KS geometry. On the point
 across the positive and negative coupling constant region, the metrics
 are continuous but non-smoothly connected, the form fields are
 discontinuous. In our studies, we assume that the negative coupling
 constant region is just as physical as the positive one, and the
 two regions are connected through a physical phase transition. One
 can also assume the existence of some physical mechanisms which
 can transmute the phase of the negative coupling
 constant into the corresponding axion field and make the
 coupling constant become a positive one. Looking for this kind of
 physical mechanisms may be an interesting topic for future works.

  \FloatBarrier

 \end{document}